\documentclass[aps,prl,twocolumn,showpacs]{revtex4}

\usepackage[utf8]{inputenc}
\usepackage{graphicx}
\usepackage{amsmath}

\begin{document}

\title{Constraint driven condensation in large fluctuations of linear statistics}

\author{Juraj Szavits-Nossan}
\email{jszavits@staffmail.ed.ac.uk}
\affiliation{SUPA, School of Physics and Astronomy, University of Edinburgh, Mayfield Road, Edinburgh EH9 3JZ, United Kingdom}

\author{Martin R. Evans}
\email{mevans@staffmail.ed.ac.uk}
\affiliation{SUPA, School of Physics and Astronomy, University of Edinburgh, Mayfield Road, Edinburgh EH9 3JZ, United Kingdom}

\author{Satya N. Majumdar}
\email{majumdar@lptms.u-psud.fr}
\affiliation{Laboratoire de Physique Th\'{e}orique et Mod\`{e}les Statistiques, UMR 8626, Universit\'{e}
Paris Sud 11 and CNRS, B\^{a}t. 100, Orsay F-91405, France}

\begin{abstract}
Condensation is the phenomenon whereby one of a sum of random variables contributes a finite fraction to the sum.  It is manifested as an aggregation phenomenon in diverse physical systems such as coalescence in granular media, jamming in traffic and gelation in networks. We show here that the same condensation scenario, which normally happens only if the underlying probability distribution has tails heavier than exponential, can occur for light-tailed distributions in the presence of additional constraints. We demonstrate this phenomenon on the sample variance, whose probability distribution conditioned on the particular value of the sample mean undergoes a phase transition. The transition is manifested by a change in behavior of the large deviation rate function.
\end{abstract}

\maketitle


In recent years the study of large deviations has come to the fore as a general framework in statistical physics \cite{Touchette09}. This has proved especially useful in the domain of far-from-equilibrium phenomena where the form of stationary states is not immediately available.  The standard large deviation theory describes the probability of events far away from the mean as $P(\Delta x)\propto\exp[-LI(\Delta x/L)]$ where $\Delta x\propto L$ is the deviation from the mean of some extensive observable $x$, $I$ is the rate function and $L$ is the system size. This contrasts with small deviations which are typically described by a Gaussian distribution. The fact that $P$ is dominated by the minimum of $I$ occurring at $\Delta x$=0 implies that large deviation rate functions may play the role of the free energy for nonequilibrium systems \cite{DL98,DLS01,Derrida07}. Similar large deviation principles also exist for time-extensive quantities such as the current or the activity in lattice gases \cite{BSGJLL05,BD05,GJLPDW09}. 

It is often observed that profound physical phenomena such as phase transitions have a striking manifestation  in the large deviation theory. For example, the nonequivalence of microcanonical and canonical ensembles found in many systems with long-range interactions is manifested by non-convexity of the rate function \cite{Touchette09}. Furthermore, nonanalytic behavior of $I$ signals a nonequilibrium phase transition, just as nonanalyticity of the free energy corresponds to an equilibrium phase transition. Recently there has been considerable interest in such nonequilibrium phase transitions \cite{BD05,BSGJLL05,GJLPDW09,Bertini10,HG11,Bunin12}. 

A simple, but significant, example is the condensation phase transition \cite{note} which occurs, for example, in classical systems of interacting particles such as the zero-range process (ZRP) or models for the transfer of a continuous mass variable (for a comprehensive review see \cite{EvansHanney05}). Remarkably, even though the dynamics contains interactions in these models, the steady-state probability of a microstate comprising masses $m_i$ at sites $i= 1, \ldots L$ takes a factorized form given by
\begin{equation}
P(m_1,\dots,m_L)=\frac{1}{Z_L}\prod_{i=1}^{L}f(m_i)\delta(M-\sum_{i}m_i),
\label{P}
\end{equation}

\noindent where $Z_L$ is the normalization constant given by
\begin{equation}
Z_L(M) =\prod_{i=1}^{L}\int {\rm d} m_i f(m_i) \delta(M-\sum_{i}m_i).
\label{Z}
\end{equation}

\noindent Here correlations between sites are only generated by the global constraint imposed by the delta function that the total particle number is fixed to be $M$. When the underlying weight $f(m)$ is heavy-tailed in a sense that its tail decays slower than exponential, a condensation phase transition occurs as the global density of particles is raised above a critical value. In the condensed phase, a randomly selected site carries a macroscopic fraction of the particles which is referred to as the condensate. The condensation transition exhibited by the ZRP has thus served as a baseline model for studying  aggregation phenomena in systems with more complex states such as  coalescence in granular systems \cite{Torok05}, jamming in traffic \cite{OEC98,KMH05}, Ostwald ripening \cite{Reis08}, gelation in networks \cite{Krapivsky00}, emulsification failure in polydisperse hard-sphere systems \cite{EMPT10} etc. The dynamics of condensation has also proved of interest \cite{G03,GL05,GSS03} and recently generalizations to moving condensates have revealed curious dynamical effects \cite{HMS09,WE12}.

The connection between the simple factorized steady state of the ZRP and large deviations of a sum of random variables can be seen when one realizes that $P(m_1,\dots,m_L)$ (Eq. \ref{P}) is equivalent to the probability of picking $L$ independent and identically distributed  random variables (iidrv) from a common distribution $f(m)$ (provided we normalize it properly), conditioned on the fixed value of their total sum \cite{MEZ05,EMZ06}. In that context the partition function $Z_L(M)$ itself becomes a probability distribution for the sum of $L$ iidrv. By fixing $M=\mu L$, where $\mu\neq \langle m\rangle$, we condition random numbers on large deviations of their sum. Thus the constraint of fixed total mass, which is imposed by the dynamics of models such as ZRP, provides an elegant way of probing what would otherwise be rare events \cite{YK}.

In this work we go further and consider the effect of several global  constraints on the simplest scenario of a factorized steady state. Our key observation is that condensation may be observed even when the underlying distribution $f(m)$ is light-tailed. Our central object of study is the following partition function
\begin{eqnarray}
Z_L(M,V)&=&\int dm_1\dots\int dm_L\prod_{i=1}^{L}f(m_1) \nonumber\\ 
&&\times \delta(M-\sum_i m_i)\delta(V-\sum_i m_{i}^{1/p}).
\label{Z_M_V}
\end{eqnarray}

\noindent Generally we consider the linear statistic $V=\sigma L=\sum_{i}m_{i}^{1/p}$, but we shall first focus on the case $p=1/2$ in which case $V-M^2$ is the sample variance of $L$ rvs
drawn from $f(m)$. In physical systems, the sample variance plays a crucial role in determining the strength of fluctuations and generally has a non-trivial probability distribution. The
sample variance is of particular significance in models of traffic flow where $m_i$ correspond to the headway between adjacent vehicles $i$ and $i+1$ and the sample variance gives a global
measure of how bunched a set of $L$ vehicles in a fixed spatial region is \cite{OEC98}. Another example is the much studied field of interface growth where $m_i$ corresponds to the height at site $i$ and $V$ then corresponds to the roughness of the interface \cite{FamilyVicsek85,Krug97,PRZ94}. Also we can mention volatility in financial models defined as the variance of random variables $m_i$ taking the form of log returns \cite{volatility}. Then $i$ indexes the time interval and $m_i$ is the log of ratio of the price of a stock at time $i$ and $i-1$. 
In all these contexts a key question is how does a large deviation in the sample variance arise. The computation of (\ref{Z_M_V}) gives information on the conditioned probability density 
$P(V\vert M)$ of finding $V=\sum_i m_{i}^{1/p}$ given the value of the sum $M=\sum_i m_i$ since one can write $P(V\vert M)$ as
\begin{equation}
P(V\vert M)=\frac{Z_L(M,V)}{Z_L(M)}.
\end{equation}

We shall show that asymptotic form of $\ln Z_L(M,V)/L$, which forms the large deviation function, changes form according to the values of $M$ and $V$ thus indicating a condensation transition.
We shall first demonstrate condensation transition for the exponential distribution $f(m)=r\exp(-rm)$ which belongs to the class of light-tailed  distributions  which have exponentially
bounded tails. After a straightforward generalization to other light-tailed cases, we shall conclude by analyzing the interesting case of heavy-tailed distributions, for which we
find that the presence of two constraints can suppress the condensation that would otherwise happen if only one of the constraints was present.


\noindent \textit{Phase diagram.} Our strategy for finding $Z_L(M,V)$ is to calculate its Laplace transform $\tilde{Z}_L(s,\lambda)$ with respect to $M$ and $V$. The biggest advantage of working with iidrvs
comes from the known form of $\tilde{Z}_L(s,\lambda)$, which has a factorizing property $\tilde{Z}_L(s,\lambda)=[g(s,\lambda)]^L$, where 
$g(s,\lambda)=\int_{0}^{\infty}dm f(m)e^{-sm-\lambda m^{1/p}}$. The partition function $Z_L(M,V)$ can be then found by the inversion formula
\begin{eqnarray}
Z_L(M,V)&=&\int_{c-i\infty}^{c+i\infty}\frac{ds}{2\pi i}\int_{d-i\infty}^{d+i\infty}\frac{d\lambda}{2\pi i}\nonumber\\
&&\times\exp\left[ sM+\lambda V+L\text{ln}g(s,\lambda)\right],\label{ZMV}
\end{eqnarray}

\noindent where the constants $c$ and $d$ are chosen to be right of any singularities. When $L$ is large, it is natural to evaluate integrals in $Z_L(M,V)$ using the saddle-point approximation, which amounts to solving the following saddle point equations for $s$ and $\lambda>0$,
\begin{subequations}
\label{saddle_point}
\begin{eqnarray}
\mu&=&\frac{\int_{0}^{\infty}m f(m)e^{-sm-\lambda m^{1/p}}dm}{g(s,\lambda)}\label{saddle_mu}\\
\sigma&=&\frac{\int_{0}^{\infty}m^{1/p} f(m)e^{-sm-\lambda m^{1/p}}dm}{g(s,\lambda)}.\label{saddle_sigma}
\end{eqnarray}
\end{subequations}

\noindent For $f(m)=r\exp(-rm)$ and $p=1/2$, the integrals in (\ref{saddle_mu}) and (\ref{saddle_sigma}) can be expressed in terms of the complementary error function $\text{erfc}(z)$,
\begin{subequations}
\label{saddle_point_1}
\begin{eqnarray}
\mu&=&\frac{1-\sqrt{\pi}z e^{z^2}\text{erfc}(z)}{\sqrt{\lambda}\sqrt{\pi}e^{z^2}\text{erfc}(z)}\equiv\frac{F_1(z)}{\sqrt{\lambda}} \label{F_1}\\
\sigma&=&\frac{(1+2z^2)e^{z^2}\text{erfc}(z)-2z/\sqrt{\pi}}{2\lambda e^{z^2}\text{erfc}(z)}\equiv\frac{F_2(z)}{2\lambda},\label{F_2}
\end{eqnarray}
\end{subequations}

\noindent where we used shorter notation $z=(s+r)/(2\sqrt{\lambda})$ and introduced functions $F_1(z)$ and $F_2(z)$. We can now combine (\ref{F_1}) and (\ref{F_2}) in a single equation for $z$ in terms of $F_1(z)$ and $F_2(z)$,
\begin{equation}
\frac{2\sigma}{\mu^2}=\frac{F_2(z)}{F_{1}^{2}(z)}.\label{saddle_z}
\end{equation}

\begin{figure}[hbtp!]
\includegraphics[width=8.6cm]{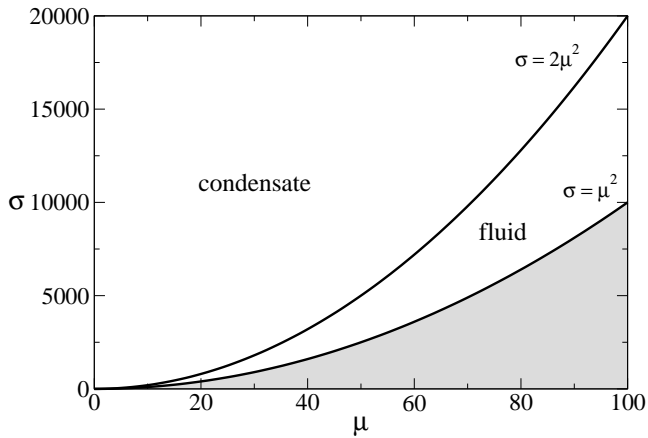}
\caption{\label{fig1} Phase diagram in the $\mu-\sigma$ plane for $f(m)=re^{-rm}$ and $p=1/2$. Shaded area $\sigma<\mu^{2}$ is forbidden owing to Jensen's inequality.}
\end{figure}

\noindent Using the known limiting behaviors of complementary error function, $\text{erfc}(z)\rightarrow\exp(-z^2)/(z\sqrt \pi)\left(1- 1/(2z^2)\right)$ as $z\rightarrow\infty$ and
$\text{erfc}(z)\rightarrow 2$ as $z\rightarrow -\infty$, we can show that (\ref{saddle_z}) admits a solution only when $\mu^2<\sigma<2\mu^2$.
The lower bound $\mu^2<\sigma$ is just Jensen's inequality which states that $V\leq M^2$, while the upper bound implies a breakdown of the saddle-point approximation for $\sigma>2\mu^2$ and signals a phase transition. We shall show below that the transition line $\sigma_c(\mu)=2\mu^2$, as presented in figure \ref{fig1}, separates a `fluid' phase ($\mu^2<\sigma<2\mu^2$) from the phase with a `condensate' ($\sigma>2\mu^2$). In the fluid phase all $L$ random variables  contribute `cooperatively' to accommodate atypical values of $M$ and $V$ by tuning $s$ and $\lambda$ that solve (\ref{F_1}) and (\ref{F_2}), respectively. On the contrary, large deviations in the condensed phase are typically realized by
the square of one of the random variables contributing a macroscopic fraction of $V$.


\noindent \textit{Canonical Analysis of The Phase Diagram.} A detailed study of the condensed phase requires a canonical approach that goes beyond saddle point analysis, i.e. we need to evaluate (\ref{ZMV}) when the saddle-point approximation breaks down. The idea is to consider $\Omega_L(s,V)$, the Laplace 
transform of $Z_L(M,V)$ with respect to $M$,
\begin{eqnarray}
\Omega_L(s,V)&=&\int_{0}^{\infty}dm_1\dots \int_{0}^{\infty}dm_L \prod_{i=1}^{L}f(m_i)e^{-sm_i}\nonumber\\
&&\times\delta\left(V-\sum_{i}m_{i}^2\right),
\end{eqnarray}

\noindent and then use some results from \cite{EMZ06,EMPT10,Nagaev69} to obtain the asymptotic behavior. Making a change of variables $v_i=m_{i}^{2}$ and defining $w(v;s)=rv^{-1/2}\text{exp}^{-(s+r)v^{1/2}}/(2g(s,0))$ so that $\int_{0}^{\infty}w(v)dv=1$, 
we can write $\Omega_L(s,V)=[g(s,0)]^L\Pi_L(V;s)$, where $\Pi_L(V;s)$ is given by
\begin{eqnarray}
\Pi_L(V;s)&=&\int_{0}^{\infty}dv_1\dots dv_L \prod_{i=1}^{L}w(v_i;s)\nonumber\\
&&\times\delta\left(V-\sum_{i}v_i\right).
\label{pi}
\end{eqnarray}

\noindent When the variable $s$ in (\ref{pi}) is real, $\Pi(s,V)$ becomes the probability density for the sum $V=\sum_i v_i$ of $L$ iidrvs having a common distribution $w(v;s)$. 
The presence of $\exp[-(r+s)v^{1/2}]$ in $w(v;s)$ makes $w(v;s)$ heavy-tailed (in this case it is a Weibull distribution), and therefore the usual condensation transition is expected 
for $V=L\sigma>V_c(s)=L\sigma_c(s)$, where $\sigma_c(s)=\langle v\rangle_w=2(r+s)^{-2}$ is the first moment with respect to $w(v;s)$ \cite{EMZ06}. For $\sigma>\sigma_c(s)$, we can invoke an 
old result of \cite{Nagaev69} for large deviations of the sum of iid random variables that follow a Weibull distribution, which states that for large $L$,
\begin{equation}
\Pi(L\sigma;s)\approx Lw(L\sigma-L\sigma_c(s);s).\label{nagaev}
\end{equation}

\begin{figure}[!hbt]
\includegraphics[width=8.6cm]{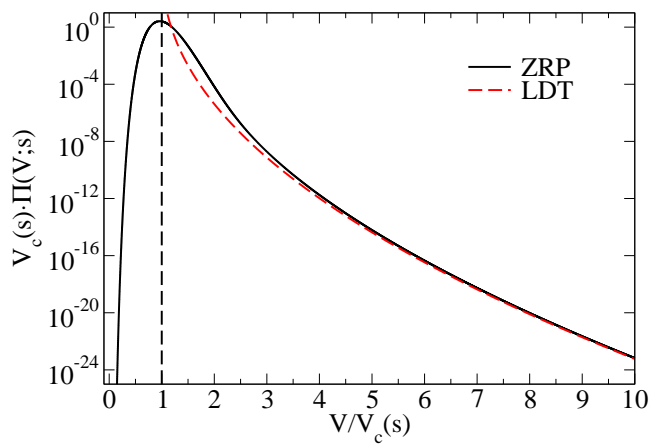}
\caption{\label{fig2} (Color online) Expression (\ref{nagaev}) from large deviation theory (LDT, dashed line) compared to $\Pi(V;s)$ (ZRP, solid line). $\Pi_L(V;s)$ was obtained numerically for $L=200$ and $r+s=-\ln(0.9)$ via recursion relation $\Pi_L(V;s)=\sum_v w(v;s)\Pi_{L-1}(V-v;s)$ valid for integer random variables \cite{EvansHanney05}, where $w(v;s)$ is discrete version of  Weibull distribution, $w(v;s)=\exp(-(r+s)\sqrt{v})-\exp(-(r+s)\sqrt{v+1})$ and $v$ is an integer. For large $v$, $w(v;s)$ reduces to the continuous Weibull distribution defined earlier.}
\end{figure}

\noindent Again, this result states that the large deviation of $V$ arises through $L-1$ random variables having typical values around $\sigma_c$ and one random variable taking the value $L(\sigma-\sigma_c(s))$ \cite{Nagaev69}. A comparison with exact $\Pi(V;s)$ obtained numerically for integer random variables, which is presented in figure \ref{fig2}, shows that (\ref{nagaev}) is a good approximation of $\Pi_L(V;s)$ even for lattices of size $L=200$.


While the above analysis pertains to real $s$, to find $Z_L(M,V)$ by the inversion formula
\begin{equation}
Z_L(M,V)=\int_{c-i\infty}^{c+i\infty}\frac{ds}{2\pi i}e^{sM+L\text{ln}g(s,0)}\Pi(V;s),\label{Z_M_V_pi}
\end{equation}

\noindent we have to allow complex $s$ for which (\ref{nagaev}) does not hold in general. However, we can calculate the integral in (\ref{Z_M_V_pi}) using the saddle-point method by 
noting that for any complex $s$ with $s_0=\text{Re}(s)$, the following inequality $\vert\Pi_L(V;s)\vert\leq \Pi_L(V;s_0)$ guarantees that $\Pi_L(V;s)$ decays slower 
than exponential in $L$. Calculating $Z_L(M,V)$ then amounts to solving  
$\mu=-\partial_s g(s,0)/g(s,0)$, which gives  $s^{*}=1/\mu-r$. A final step is to recognize that the integral we are then left with is in fact the saddle-point approximation of $Z_L(M)$, so that $Z_L(M,V)\approx L\,w(L\sigma-L\sigma_c;s^{*})Z_L(M)$, where $\sigma_c=\sigma_c(s^{*})=2\mu^2$. The final result  for $\sigma>2\mu^2$ and large $L$ is  
\begin{equation}
P(V=L\sigma\vert M=L\mu)\approx\frac{\sqrt{L}e^{-\mu^{-1}\sqrt{L(\sigma-2\mu^2)}}}{2\mu\sqrt{\sigma-2\mu^2}}.
\end{equation}

\noindent This result is clearly different from what we find in the fluid phase, where $P(V\vert M)$ decays exponentially fast in $L$ with rate function $I(\mu,\sigma)=s^*\mu+\lambda^*\sigma+\textrm{ln}g(s^*,\lambda^*)-\textrm{ln}\mu-1$, where $s^*$ and $\lambda^*$ solve (\ref{saddle_mu}) and (\ref{saddle_sigma}). The condensation transition thus marks 
a change in the rate at which $P(V\vert M)$ decays for large $L$ that goes from exponential in the fluid phase to subexponential in the condensed phase. [In a finite system, this change 
of scale however does not happen exactly for $\sigma=\sigma_c$. Close to the transition point where $\vert\sigma-\sigma_c\vert\sim O(L^{-1/2})$, one can show that $P(V\vert M)$ is smooth 
and follows Gaussian distribution with the standard deviation $2\mu^2\sqrt{L}$. The transition line is thus shifted by the amount $4\mu^2 L^{-1/3}$ \cite{SEM}.]


\noindent \textit{Generalizations.} So far we have considered the exponential distribution $f(m)=r\exp(-rm)$ and $p=1/2$. Most generally, we have to distinguish between distributions that decay exponentially or faster (light-tailed) or slower than exponential (heavy-tailed). For simplicity, we will consider here only distributions that either have strictly exponential tails, $f(m)\sim e^{-rm}$ for large $m$, or tails that are not exponentially bounded (heavy-tailed distributions). We also have to distinguish between cases $p<1$ and $p>1$. The fundamental difference in going from $p<1$ to $p>1$ is that for $p>1$ the condensation transition happens whereby one of the $m_i$'s takes a macroscopic fraction of $M$ rather than $V$. Here, we consider only the $p<1$ case and leave $p>1$ for a subsequent publication \cite{SEM}.

For a general distribution $f(m)$ and $p \neq 1/2$  we can no longer perform the integrals in (\ref{saddle_mu}) and (\ref{saddle_sigma}) explicitly. However, by 
considering the r.h.s. of (\ref{saddle_mu}) and (\ref{saddle_sigma}) as functions $\mu(s,\lambda)$ and $\sigma(s,\lambda)$ respectively, we can show that 
 (a) $\mu=\mu(s,\lambda)$ has a unique solution $s_{\mu}(\lambda)$ for any given $\mu$ and any $\lambda>0$, and (b) $\sigma=\sigma(s_{\mu}(\lambda),\lambda)$ has a unique solution 
provided $\sigma<\sigma(s_{\mu}(0),0)$. The limiting value $\sigma_c(\mu)=\sigma(s_\mu(0),0)$ arises because $\lambda$ must always stay non-negative, otherwise the integrals in 
(\ref{saddle_mu}) and (\ref{saddle_sigma}) will diverge. Proofs of (a) and (b) are straightforward \cite{SEM}.

\begin{figure}[!hbt]
\includegraphics[width=8.6cm]{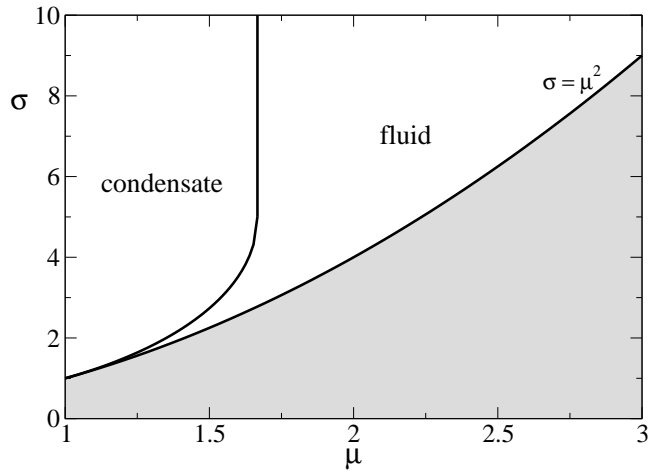}
\caption{\label{fig3} Phase diagram in the $\mu-\sigma$ plane for $p=1/2$ and the Pareto distribution $f(m)=(\gamma-1)/m^{\gamma}$, $m\geq 1$, where $\gamma=7/2$. Shaded area $\sigma<\mu^{2}$ is forbidden owing to Jensen's inequality.}
\end{figure}

The question then arises as to what happens for $\sigma>\sigma_c(\mu)$? Notice that so far we have not distinguished  between light-tailed and heavy-tailed distributions, which becomes important when considering the limit $\text{lim}_{\lambda\rightarrow 0}\sigma(s_\mu(\lambda),\lambda)$. If this limit is finite for any $\mu$, then $\sigma_c(\mu)=\sigma(s_\mu(0),0)$ becomes the transition line separating fluid from condensed phase and the corresponding phase diagram will be as in figure \ref{fig1}. This type of behavior will happen for any  light-tailed distribution with a tail $\propto\exp(-rm)$, in which case $s_\mu(0)>-r$ can solve (\ref{saddle_mu}) for any $\mu$. On the other hand, if $f(m)$ has a tail that decays slower than exponential, then $s_\mu(0)$ does not exist for $\mu>\mu_c$, nor does the transition line $\sigma_c(s_\mu(0),0)$. Instead, both saddle point equations can be solved with some positive $\lambda$ and negative $s_\mu(\lambda)$. The corresponding phase diagram, presented in figure \ref{fig3} for a Pareto distribution, has a transition line that follows $\sigma_c(\mu)=\sigma(s_\mu(0),0)$ for $\mu<\mu_c$ and then turns into straight line $\mu=\mu_c$ for $\sigma>\sigma_c(\mu_c)=\langle m^{1/p}\rangle$. For $\mu>\mu_c$, it is thus no longer possible to go to the condensed phase by increasing $\sigma$. For $\mu>\mu_c$ there is also an interesting effect that fixing $V=\sigma L$ has suppressed condensation that would have appeared in $M$ had there been no constraint on $V$.

%


To summarize, we have shown how several constraints on the steady
states of stochastic systems can lead to condensation phenomena even
when the underlying distribution is light-tailed. We have analyzed
this effect in detail in the simplest physical scenario of a
factorized steady state and constrained sample variance, which reduces
to computing large deviation rate functions of sums of independent
random variables.  However, the phenomenon should extend to other
linear statistics and even to correlated random variables. An
interesting example of the latter case is the Renyi entropy used to
describe quantum entanglement. To compute the distribution of this
entropy one has to consider the distribution of eigenvalues of reduced
density matrix, which are correlated random variables, subject to
constraints. Recent results have shown that condensation-like
transitions in the eigenvalue distribution can occur
\cite{Nadal10,Nadal11}. Finally, to put our findings in the broader
context, we mention recent results on current fluctuations in
diffusive systems \cite{BD05,BSGJLL05,HG11} where it was shown how
large fluctuations beyond some critical current are realized in a very
specific, organized way that resembles a condensate. 

\textit{Note Added.} After we submitted this work, we received a manuscript \cite{Filiasi13} which deals with a related problem and recovers figure \ref{fig3}. Also we became aware of a close connection of condensation in our work to breathers (localized solutions) in the discrete nonlinear Schr\"{o}dinger equation and related models \cite{Iubini13}. Finally, the $p=1/2$ case with a uniform underlying distribution has been studied using probabilistic techniques in \cite{Chatterjee10}.

\begin{acknowledgments}
JSN and MRE would like to acknowledge funding from EPSRC under grant number EP/J007404/1. SNM acknowledges support by ANR grant 2011-BS04-013-01 WALKMAT. 
\end{acknowledgments}



\begin{thebibliography}{99}
\bibitem{Touchette09} H. Touchette, \textit{Phys. Rep.} \textbf{478}, 1-69 (2009)
\bibitem{DL98} B. Derrida and J. L. Lebowitz, \textit{Phys. Rev. Lett.} \textbf{80}, 209 (1998)
\bibitem{DLS01} B. Derrida, J. L. Lebowitz and E. R. Speer, \textit{Phys. Rev. Lett.} \textbf{87}, 150601 (2001)
\bibitem{Derrida07} B. Derrida, \textit{J. Stat. Mech.} P07023 (2007)
\bibitem{BSGJLL05} L. Bertini, A. De Sole, D. Gabrielli, G. Jona-Lasinio and C.Landim, \textit{Phys. Rev. Lett.} \textbf{94}, 030601 (2005)
\bibitem{BD05} T. Bodineau and B. Derrida, \textit{Phys. Rev. E} \textbf{72}, 066110 (2005)
\bibitem{GJLPDW09} J.P. Garrahan, R. L. Jack, V. Lecomte, E. Pitard, K. van Duijvendijk and F. van Wijland, \textit{J. Phys. A: Math. Theor.} \textbf{42}, 075007  (2009)
\bibitem{Bertini10} L. Bertini, A. De Sole, D. Gabrielli, G. Jona-Lasinio and C. Landim, \textit{J. Stat. Mech.} L11001 (2010)
\bibitem{HG11} P.I. Hurtado and P.L. Garrido, \textit{Phys. Rev. Lett.} \textbf{107}, 180601 (2011)
\bibitem{Bunin12} G. Bunin, Y. Kafri and D. Podolsky, \textit{J. Stat. Mech.} L10001 (2012)
\bibitem{note} This form of condensation bears a close relation to Bose-Einstein condensation in the ideal Bose gas but is distinct from the liquid-gas transition which is also sometimes referred to as condensation.
\bibitem{EvansHanney05} M. R. Evans and T. Hanney, \textit{J. Phys. A: Math. Gen.} \textbf{38}, R195 (2005)
\bibitem{Torok05} J. T\"{o}r\"{o}k, \textit{Physica A} \textbf{355} 374-382 (2005)
\bibitem{OEC98} O.J. O'Loan, M. R. Evans and M. E. Cates \textit{Phys. Rev. E} \textbf{58}, 1404-1418 (1998) 
\bibitem{KMH05} J. Kaupu\v{z}s, R. Mahnke,  R. J. Harris, \textit{Phys. Rev. E}  \textbf{72}, 056125 (2005) 
\bibitem{Reis08} F. D. A. Aar\~{a}o Reis and R. B. Stinchcombe, \textit{Phys. Rev. E} \textbf{77}, 041411 (2008)
\bibitem{Krapivsky00} P. L. Krapivsky, S. Redner and F. Leyvraz, \textit{Phys. Rev. Lett} \textbf{85}, 4629-4632 (2000) 
\bibitem{EMPT10} M. R. Evans, S. N. Majumdar, I. Pagonabarraga,  E. Trizac, \textit{J. Chem. Phys.} \textbf{132}, 014102 (2010)
\bibitem{G03} C. Godr\`{e}che, \textit{J. Phys. A: Math. Gen.} \textbf{36} 6313 (2003)
\bibitem{GL05} C. Godr\`{e}che and J.-M. Luck, \textit{J. Phys. A: Math. Gen.} \textbf{38}, 7215 (2005)
\bibitem{GSS03} S. Grosskinsky, G. M. Sch\"{u}tz, H. Spohn, \textit{J. Stat. Phys.} \textbf{113},  389-410 (2003)
\bibitem{HMS09} O. Hirschberg, D. Mukamel, and G. M. Sch\"{u}tz, \textit{Phys. Rev. Lett.} \textbf{103}, 090602 (2009) 
\bibitem{WE12} B. Waclaw and M.  R. Evans, \textit{Phys. Rev. Lett.} \textbf{108}, 070601 (2012)
\bibitem{MEZ05} S. N. Majumdar, M. R. Evans and R. K. P. Zia, \textit{Phys. Rev. Lett.} \textbf{94}, 180601 (2005)
\bibitem{EMZ06} M. R. Evans, S. N. Majumdar and R. K. P. Zia, \textit{J. Stat. Phys} \textbf{123}, 357-90 (2006)
\bibitem{YK} N. Merhav and Y. Kafri, \textit{J. Stat. Mech} P02011 (2010)
\bibitem{FamilyVicsek85}F. Family and T. Vicsek,  \textit{J. Phys. A: Math. Gen.} \textbf{18}, L75 (1985)
\bibitem{Krug97} J. Krug, \textit{Adv. Phys.} \textbf{46}(2) 139-282 (1997)
\bibitem{PRZ94} M. Plischke, Z. R\'{a}cz and R. K. P. Zia, \textit{Phys. Rev. E} \textbf{50}, 3589-3593 (1994)
\bibitem{volatility} Whether the stock prices are true iidrvs is still much debated today.
\bibitem{Nagaev69} A. V. Nagaev, \textit{Theory Probab. Appl.} \textbf{14}(1), 51-64 (1969)
\bibitem{SEM} J. Szavits Nossan, M. R. Evans and S. N. Majumdar, \textit{in preparation}
\bibitem{Nadal10} C. Nadal, S.N. Majumdar and M. Vergassola, \textit{Phys. Rev. Lett.} \textbf{104}, 110501 (2010)
\bibitem{Nadal11} C. Nadal, S.N. Majumdar and M. Vergassola, \textit{J. Stat. Phys.} \textbf{142}, 403-438 (2011)
\bibitem{Filiasi13} M. Filiasi, E. Zarinelli, E. Vesselli and M. Marsili, arXiv:1309.7795
\bibitem{Iubini13} S. Iubini, A. Politi and P. Politi, arXiv:1308.4870
\bibitem{Chatterjee10} S. Chatterjee, arXiv:1011.4043
\end{thebibliography}
\end{document}